# Control of Length and Spatial Functionality of Single-Wall Carbon Nanotube AFM Nanoprobes

*Haoyan Wei†, Sang Nyon Kim‡, Minhua Zhao†, Sang-Yong Ju‡, Bryan D. Huey†, Harris L. Marcus†\* and Fotios Papadimitrakopoulos‡\**

† Materials Science and Engineering Program, Department of Chemical, Materials and Biomolecular Engineering, Institute of Materials Science, University of Connecticut, Storrs, CT 06269, USA

‡ Nanomaterials Optoelectronics Laboratory, Polymer Program, Institute of Materials Science, Department of Chemistry, University of Connecticut, Storrs, CT 06269, USA

Email Contact: Haoyan Wei, wtaurus@msn.com

ABSTRACT

Single-wall carbon nanotube (SWNT) nanofibrils were assembled onto conductive atomic force microscopy (AFM) probes with the help of dielectrophoresis (DEP). This process involved the application of a 10V, 2 MHz, AC bias between a metal-coated AFM probe and a dilute suspension of SWNTs. This exerted a positive dielectrophoretic force onto the nanotubes that caused them to align while precipitating out onto the probe. The gradual removal of the AFM probe away from the SWNT suspension consolidated these nanotubes into nanofibrils with a high degree of alignment as demonstrated with polarization Raman experiments. By varying the pulling speed, immersion time and concentration of the SWNT suspension, one can tailor the diameter, and thus the stiffness of these probes. Precise length trimming of these nanofibrils was also performed by their gradual immersion and dissolution into a liquid that strongly interacted with nanotubes, (*i.e.* sodium dodecyl sulfate (SDS) solution). Vacuum annealing these nanoprobes at temperature up to 450°C further increased their stiffness and rendered them insoluble to SDS and all other aqueous media. Re-growth of a new SWNT nanofibril from the side or at the end of a previously grown SWNT nanofibril was also demonstrated by a repeated dielectrophoretic assembly at the desired immersion depth. These SWNT nanofibril-equipped AFM probes are electrically conductive and mechanically robust for use as high-aspect-ratio electrochemical nanoprobes.

# 1. Introduction

Atomic force microscopy (AFM) is a powerful analytical method capable of simultaneous surface characterization of various topological, chemical, mechanical and electrical properties.[1-3] The ability to function in a variety of different environments (*i.e.* vacuum, ambient, fluid, etc.) further increases its versatility. The spatial resolution and depth profiling of AFM imaging are typically controlled by tip radius sharpness and aspect ratio.[4, 5] Normal AFM probes are usually made from Si or $Si_3N_4$ and have pyramidal shapes with cone angles of 20-30 degrees.[6] In spite of continuous efforts to increase aspect ratio and reduce tip radius, tip-to-tip variation[6] and materials wear[7] make it difficult to further improve cone-shaped AFM probes. The high aspect ratio, small diameter, high stiffness and buckle flexing capabilities of carbon nanotubes (CNTs) render them ideal for the development of next generation AFM probes.[8, 4, 7, 9] These probes can find a number of unique applications in localized electrochemistry[10, 11] within cellular environments,[6] where membrane piercing is necessary.[12]

A number of methods have been investigated to either attach or directly grow CNTs on the apex of AFM probes. Dai *et al.*,[13] reported the initial fabrication of a multi-wall CNT (MWNT) probe by sticking a small MWNT bundle onto a cone-shaped tip, coated with adhesive. Alternatively, pre-grown MWNTs were electrostatically (DC field) oriented and attached onto AFM probes.[14-16] Using a pick-up approach, Lieber *et al.*[17] successfully reduced tip radius by realizing a single-wall CNT (SWNT) on an AFM tip. In parallel, chemical vapor deposition (CVD) based growth methods have been developed to improve reproducibility and scale up production.[5, 18] Latest studies, however, have shown relatively large length variation for CVD-grown nanotube probes.[19] The polarizable nature of CNTs has recently enabled their dielectrophoretic (DEP) assembly towards the high field region of the conductive apex, producing CNT nanofibrils with tunable length and diameter.[20-22] Unlike individual nanotubes, nanofibrils containing multiple nanotubes provide a better control over probe stiffness at high aspect ratios. In addition, parallel fabrication has also been demonstrated by the simultaneous pulling from multiple scanning tunneling microscopy (STM) tips.[23]

Despite the already achieved advances in dielectrophoretic assembly, a number of limitations remain with respect to fine tuning fibril length, diameter, stiffness as well as improving base adhesion at the AFM apex. In this paper, we provide a comprehensive investigation on growth conditions, annealing temperature and nanotube orientation that ultimately control the above mentioned fibril properties. We also introduce a novel, surfactant-assisted, dissolution process to further optimize control of the length and stiffness of these SWNT fibrils. In addition, radial and longitudinal heterostructures have been realized by sequential DEP assembly on the side walls and ends of previously grown and annealed nanofibrils, respectively.

## 2. Experimental Section

Nitric acid (98%) and sulfuric acid (96.4%) were obtained from Aldrich and used as is. A.C.S. reagent dimethylformamide (DMF) was purchased from J.T.Baker. Millipore quality deionized water with resistivity larger than 18MΩ was used for all experiments. Sodium dodecyl sulfate (SDS, ≥98.5%) was purchased from Sigma-Aldrich and dissolved into deionized water to produce 1wt% solution. Laser ablated SWNTs were purchased from Tubes@Rice. Following the previously established protocol,[24-27] pristine SWNTs were treated in 3:1 mixture of $H_2SO_4$ and $HNO_3$ with sonication for 4 hours at 70°C, filtered, washed with copious deionized water until the pH of filtrated water reached neutral and dried overnight in vacuum. These purified shortened-SWNTs (s-SWNTs) were dispersed in deionized water or DMF with sonication and further diluted to make nanotube solutions with different concentrations (0.001~0.01 mg/ml) at which the solution became colorless.

Commercial AFM probes (Olympus AC240 series) purchased from Asylum Research were used for carbon nanotube pulling. AC240TS are standard cantilevers for AC mode (tapping mode) imaging, with Al coated on the reflex side. AC240TM are electri-levers which have an additional Pt coating on the tip side besides the Al coating on the reflex side. Both tips are made from Si materials and have a tetrahedral tip shape with typical height of 14 μm. The spring constant is *ca.* 2 N/m, and resonant frequency is *ca.* 70 kHz. Representative images of these AFM probes from different viewing angle are provided in Figure S1 of the Supporting Information.

The experimental set-up for fabricating SWNT nanofibril-equipped AFM probes is schematically illustrated in Scheme 1, while pictures of the actual apparatus are shown in Figure S2 of the Supporting Information. An AFM probe was taken as the working electrode and placed on a conductive platform with its surface tilted between 0° and 35° with respect to the horizontal plane of the platform. A small hollow metal tube (outer diameter 600 μm, inner diameter 150 μm) was used as the counter electrode. A drop of SWNT dispersion was placed into the metal tube to form a droplet at the end of the tube, while AC voltage (2 MHz, 10V) was applied between the two electrodes. In order to slow down the evaporation of SWNT solution, both AFM probes and metal tube were enclosed in a sealed environmental-cell (E-cell) equipped with a thin transparent glass slide on top, from where an optical microscope provides real-time monitoring. To obtain better imaging contrast of the AFM tips and the meniscus of the liquid drop, bottom illumination was adopted. The optical microscope was equipped with a charge-coupled device (CCD) camera connected to a computer station for monitoring and recording. The metal tube containing the nanotube solution was connected to a XYZ 3D moving stage. The Y and Z axes were manually controlled to provide alignment of the AFM tips with the summit of solution droplet at the end of the metal tube. The X axis was motorized for carbon nanotube drawing with the capability to deliver minimum incremental displacement of 50 nm.

The morphology of the SWNT nanofibrils along with their dimensions (length and diameter) was obtained from high resolution scanning electron microscopy (SEM) (JOEL 6335F field emission SEM (FESEM) equipped with a cold field emission source). The base pressure of the sample chamber was maintained at *ca.* $10^{-6}$ torr, while the accelerating voltage varied from 5~10 keV and the working distance was set between 8~15 mm.

Polarization Raman spectra of the SWNT nanofibrils were obtained on a Renishaw Ramanscope 2000 using a 785 nm laser focused on a 1 μm spot by a 50 X objective lens, at an incident laser beam power of *ca.* 3.5 mW. The SWNT nanofibrils were interrogated in both parallel and perpendicular orientation with respect to the laser polarization direction.

**3. Results and Discussion**

**3.1 Dielectrophoretic Assembly of SWNT Nanofibrils onto AFM Probes**

Dielectrophoresis (DEP) has been widely used for assembling colloidal particles[28-30] and manipulating biology entities (such as DNA, proteins and cells)[31-33] on a variety of substrates. The high-aspect-ratio and polarizable nature of SWNTs have been used to align and assemble them onto conductive scanning probes.[20-23] This is based on the motion of these dielectric objects as a result of polarization induced by a non-uniform AC electric field. As shown in Scheme 1 the sharp AFM probe versus the relatively large metal tube with the SWNT solution droplet produce a heterogeneous electrical field under an applied AC bias. This results in polarized nanotube rotation to align their long dimension parallel to the electric field.[20, 22] The torque exerted on the nanotubes is a function of the applied electric field,[34] and for the 10V, 2 MHz AC bias used in this study, the estimated value is on the order of $4 \times 10^{-19}$ N m.[23] Such torque can generate a force of *ca.* $10^{-13}$ N at the end of a micro long nanotube, which is significantly larger than the gravitational force (on the order of $10^{-19}$ N)[34] and Brownian motion[35] imparted on the nanotube. Besides torque, the polarized nanotubes are also experiencing the anisotropy of the electric field that causes a translational movement toward the high field region. Although SWNTs were also subject to other forces including gravity and Brownian motion, we focused our attention on DEP force only since both experimental[35] and simulation[34] results indicated that the aforementioned forces were small in comparison with the DEP force (on the order of $10^{-14}$-$10^{-10}$ N),[34] and therefore negligible. The dielectrophoretic force imparted on carbon nanotubes is given by the equation:[22]

$$F_{DEP} = 2\pi a^3 \varepsilon_m \operatorname{Re}\left[\frac{\varepsilon_p^* - \varepsilon_m^*}{\varepsilon_p^* + 2\varepsilon_m^*}\right] \nabla |E|^2 \qquad (1)$$

where $a$ is the length of carbon nanotube bundles, $\varepsilon_m$ is the dielectric constant of the medium, $\varepsilon_p$ is the dielectric constant of the particle (in this study the nanotube is the particle), $\varepsilon^*$ is the complex permittivity expression, and $E$ is the electric field. The magnitude of the force depends on not only the voltage but also the frequency of the applied electrical field as well, since the permittivity $\varepsilon^*$ is frequency-dependent based on the following complex relationship:[22, 34]

$$\varepsilon^* = \varepsilon - i\frac{\sigma}{\omega} \qquad (2)$$

where $\sigma$ is the conductivity and $\omega$ is the frequency of the electrical field. As indicated in Equation (1), in addition to the applied external electrical field $E$, the cube of the length of carbon nanotubes strongly affects the magnitude of the induced DEP force. This implies that significantly larger DEP forces are applied to longer nanotubes that are preferentially attracted to the conductive AFM tip that corresponds to the high field region.

In our experiment, the applied frequency was set at the previously reported value 2 MHz[20, 21, 23] and under this condition the assumption $\varepsilon(2\ \text{MHz}) \approx \varepsilon(0)$ could be adopted.[36, 37] Therefore the direction of the DEP force depends solely on the relative dielectric properties of carbon nanotubes with respect to the surrounding medium. In this study, deionized water and DMF, whose dielectric constants were *ca.* 80 and 39 respectively, were chosen as the nanotube dispersion medium. SWNTs as produced contained mixtures of metallic (*met-*) and semiconducting (*sem-*) tubes. *Sem*-SWNTs have finite dielectric constant with $\varepsilon_{sem} < 5$ while *met*-SWNTs are expected to have a very large value ($\varepsilon_{met}$) owing to the mobile carriers.[36] However, SWNTs are also subject to aggregation in bundles due to strong hydrophobic interaction between adjacent nanotube side walls. Since the *met*-SWNTs in the bundle will dominate the dielectric constant of the bundle,[36] the effective dielectric constant for nanotubes can thus be considered significantly larger than the surrounding medium, causing them to migrate toward the high field region (AFM tip ends) under the electric field gradient. This constitutes positive dielectrophoresis as opposed to negative dielectrophoresis, which moves particles to the low field region as a result of their relatively lower dielectric constant than the surrounding medium.

Figure 1 schematically depicts the pulling process of assembling SWNT nanofibrils onto AFM probes. This assembly process can be divided into four stages in term of pulling speed control. Initially, the SWNT solution bubble is brought closer to the AFM probe in a deceleration mode to minimize the tip wetting (Stage I). Once the AFM tip starts touching the bubble, the liquid wets the AFM tip at *ca.* 1/3 of its height (Stage II). As discussed above, the carbon nanotube bundles align themselves along the electric field direction and migrate toward the immersed AFM tip end under the DEP force. The nanotube bundles that arrive earliest deposit onto the AFM tip surface and the subsequent nanotube bundles continue to deposit on their predecessors. As the translator moves away from the AFM tip, the solution bubble dewets the AFM probe and only the grown nanofibril keeps contact with the solution (Stage III). The nanotube alignment and adhesion will be further enhanced by the compressive capillary force generated from the meniscus of the solution bubble. As the pulling was in an acceleration mode, the nanotube bundle tip grew in a conical shape with a sharper end. When the desired nanofibril length was reached, the translation was rapidly increased to a rate that the deposition of carbon nanotubes could not be maintained resulting in the SWNT nanofibril being pulled out of the solution bubble and the growth stopped (Stage IV). Except during the initial and final stages of wetting and dewetting of the AFM probe, the moving rate of the meniscus of the solution bubble was consistent with the translator moving speed, assuming the change of meniscus caused by solution evaporation was negligible.

**3.2 Parameter Optimization of SWNT Nanofibril Growth**

Figure 2 provides representative FESEM images of a SWNT nanofibril dielectrophoretically grown onto the end of an AFM probe. The typical length of SWNT nanofibril is in the range of 1~10 μm. Typically the nanofibril diameter is on the order of 100 nm (Figure 2b) at the interface with the AFM probe and gradually shrinks down to 20~30 nm at the end of the tip (Figure 2d). This conical geometry is attributed to the accelerated translation during the tip drawing (Stage III in Figure 1). As the pulling speed increases, fewer nanotubes deposit, leading to thinner fibrils with sharper ends, a method which could be utilized to fabricate sharp high spatial resolution probes. As indicated in Figure 2b, c, SWNT bundles get affixed around the AFM probe surface with tube length directions parallel to the pulling direction thereby ensuring large contact surface and strong interfacial interaction between the grown nanofibril and the AFM probe.[15, 16]

The alignment of SWNTs within the dielectrophoretically grown nanotube fibrils was investigated with polarized Raman spectroscopy. It has been previously reported that maximum intensity of CNTs is obtained when the polarization of the incident laser is parallel to the nanotube axis.[38, 39] As illustrated in Figure 3, the geometry with polarization parallel to the nanoprobe axis, results in maximum intensity. By rotating the nanoprobe 90° with respect to laser polarization direction, a dramatic decrease in intensity for both G band (1592 cm$^{-1}$) and radial breathing mode (RBM) band was observed. The peak height ratio $I_0/I_{90}$ measured from G band is *ca.* 4.8, representative of a high degree of SWNT alignment along the nanofibril direction.

These high-aspect-ratio SWNT-nanofibril probes were demonstrated to be quite suitable for imaging deep structures. Figure 4 provides the comparison of line profiles measured with normal AC240 AFM probes and SWNT-nanofibril equipped AFM probes of a Si grid (μmasch TGZ3 with nominal height of 542±2 nm), while the corresponding 3D AFM images are provided in Figure S3 of the Supporting Information. The asymmetry with the AC240 AFM probe (titling angle: 80° on the right *versus* 86° on the left) of Figure 4a originates from the anisotropic shape of this tip that is unable to reach deep in the trench from one side. In contrast, the SWNT-nanofibril probes provides a more accurate reproduction of the real profile with improved tracking as indicated in Figure 4b (tilting angle: 85° on the right *versus* 86° on the left). Electrical properties of SWNT nanofibrils were characterized by measuring *I-V* curves with the SWNT-nanofibril probe contacting a Pt substrate. The obtained *I-V* curve, Figure S4 of the Supporting Information, is approximately symmetrical with a calculated resistance of approximately 11 kΩ, indicating that these SWNT-nanofibrils are electrically conductive. This will make them useful as for point sensing of localized fields.

The nanofibril dimension (diameter and length) as well as morphology (straightness and orientation) depends on several parameters including external electric field, concentration of SWNT dispersion, immersion time, pulling rate, humidity, AFM tip wetting properties and tip alignment. Lee *et al.*[22] studied the electric conditions influencing nanotube dielectrophoretic assembly. Their findings suggest that 5V AC or larger with frequency between 100 kHz and 10 MHz is required in order to successfully form SWNT nanofibrils onto AFM probes. In this study, a 10 V, 2 MHz (peak-to-peak) AC field was adopted.[20, 21, 23] The SWNT solution used was highly diluted to the point that its color was not perceptible. Tang *et al.*[23] proposed a concentration of 0.01 mg/ml for the SWNT dispersion. In practice, we found this concentration was still too high, resulting in large aggregation of SWNT bundles. Within concentration range of 0.001~0.005 mg/ml, we could form good SWNT tips reliably. Below 0.001 mg/ml, the ability to pull continuous nanotubes dropped dramatically. A nanotube dispersion with concentration of 0.003 mg/ml was primarily utilized in our experiments. The immersion time of AFM probes in the nanotube solution is related to the solution concentration. This time could be considered as the time required for nanotubes to be transported to the apex of the AFM probe. Thicker solution requires less immersion time. However, thinner solutions are preferred since carbon nanotubes are better dispersed and fewer impurities are present. Using the above concentration, the normal immersion time is in the range of 3~10s. For less than 3s immersion time, the ability to grow SWNT nanofibrils is greatly reduced.

As mentioned previously in Section 3.1, if the pulling rate is maintained slower than the nanotube deposition rate, SWNTs grow a long and continuous nanofibril. As indicated in Figure 1, by gradually increasing the pulling rate (upwards slope in stage III), a number of special attributes can be imparted to this nanofibril drawing process: (1) The initial slow pulling rate minimizes adverse dewetting effects (initial spike in Stage III) that could lead to abrupt acceleration causing severe or transient thinning of the nanofibril that could ultimately result in undesired bending at the thinned portion, (2) The low initial pulling speed causes more nanotubes to be deposited near the AFM probe resulting in thicker base that imparts the nanofibril better strength, and (3) The gradual increase in pulling speed results in a conical shaped geometry with taper ends, rendering both good stiffness and high spatial resolution. Once the desired length was reached, the pulling speed was rapidly increased to a value exceeding the nanotube deposition rate to stop the nanofibril growth process upon separation from the solution bubble. The typical rate to halt nanofibril growth in this study is 40-60 μm/s. Due to the wide length distribution of SWNTs in the dispersion (from tens to hundreds of nanometers),[40] the nanofibril length typically varies from 1 to 10 μm, with the majority falling between 4-6 μm. This variation is also affected by AFM tip wetting variability. Typically, about one-third of the tip height is wetted for the Olympus AC240 AFM probes although considerable variations occur from probe to probe. The less the tip wetting, the better and more uniform nanofibrils are. A more detailed study controlling both SWNT length (via length fractionation) and AFM tip wetting characteristics is underway to obtain better length control of these nanofibrils. In this contribution, however, we have resolved this issue by developing an innovative method using post-growth SDS solubilization described in the next section.

Aside from SWNT length variations, the chamber humidity and AFM tip alignment also play a critical role to the quality, orientation and straightness of these nanofibrils. Since SWNTs were dispersed in deionized water or DMF, that both evaporate rapidly in an open air environment, such action causes the position of the solution meniscus to recede rapidly, as well as contaminate its liquid-air interface with uncontrollable nanotube deposits. For this, the meniscus traveling distance was utilized to quantify the evaporation rate. Figure S5a of the Supporting Information plots the meniscus moving distance against time for water dispersed SWNTs in open laboratory environment. The linear fit of the experimental data indicates an estimated moving rate of *ca.* 1.5 μm/s, clearly this speed is comparable to the initial pulling speed in Stage III (*ca.* 1-2 μm/s) rendering the entire process uncontrollable. In order to reduce the solution evaporation speed, both working and counter electrodes were enclosed in a sealed environmental cell, which was saturated with water vapors by the addition of a number of water droplets around the SWNT suspension bubble. As shown in Figure S5b of the Supporting Information, this slows the meniscus moving rate down to 18 nm/s, nearly 80 times less than in the open air environment. This improvement allows better control over the dimension of the fabricated SWNT nanofibrils using speed on the order of μm/s.

The ideal orientation for the SWNT nanofibril is parallel to the central axis of the AFM cone as shown in Figure 2a. Since these nanofibrils grow along the electric field direction, the central axis of the AFM should be parallel to this direction. Because the AC240 AFM probes used in this study were asymmetrical in geometry, supporting these tips on a flat substrate perpendicular to the drawing direction causes a misalignment of the central axis of AFM cone with the electric field direction as shown in Figure 5a. In order to achieve nanofibril alignment to the central axis of the AFM cone, these probes were mounted on a platform with a tilting angle of 26°, as shown in Figure 5b. The 26° tilt provides a number of salient benefits to the growth of these nanofibrils: (1) Focuses the electrical field to the end of the tip and removes unwanted contribution from the nearby large area conductive cantilever and its support that otherwise (at 0° tilt) defocus the field by their physical presence, (2) The central axis of the AFM cone symmetrically wets the solution bubble, thus producing a more uniform dewetting during pulling. This produces a more uniform drying force that results in a straighter and well-aligned SWNT nanofibril.

**3.3 Control of Length and Spatial Functionality of SWNT-Nanofibril AFM Probes**

**3.3.1 Shortening of SWNT Nanofibrils**

The production of high resolution electrochemical probes requires precise control over nanofibril length[11] as well as optional passivation along the sidewalls[11, 41] leaving only the nanofibril apex exposed. As explained above, although the nanofibril length could possibly be controlled during the pulling process in micrometer range, obtaining nanofibril around 1-2 μm is extremely challenging due to the initial wetting and dewetting process. In addition, longer than 4-5 μm nanofibrils are susceptible to bending or even curling due to the large internal stresses associated with drying. This limits the reliability of this method and necessitates post-fabrication shortening to remedy this shortcoming.[21] Focused ion beam (FIB),[42] electron beam[43] and electrical pulse shortening[17] have been previously employed. These techniques require sophisticated facilities, vacuum environments, and are in general time consuming. Presently, an alternative shortening methodology has been developed based on a surfactant-assisted nanofibril dissolution process. The success of this gradual dissolution, schematically shown in Figure 6, stems from the strong interaction of the surfactant with the hydrophobic sidewalls of SWNTs as well as the limited wetting of the SWNT nanofibrils due to their hydrophobic nature. Due to its amphiphilicity, sodium dodecyl sulfate (SDS) is capable to solubilize SWNTs in aqueous environments and 1 wt% SDS dispersion is capable to prevent SWNTs aggregation into bundles.[44-46] The critical micelle concentration (CMC) for SDS in water is 0.008 M[47] (*ca.* 0.25 wt%) and concentration higher than CMC tends to organize them into micellar configuration with their hydrophobic lipid chains buried inside and hydrophilic heads located on the surface. The hydrophobic cores shows preferential interaction with the sidewalls of SWNTs leading to nanotube encapsulated micelles.[44, 45] Such SDS encapsulated nanotubes are micellarly stabilized via a negative charge on their surface which prevents aggregation. This is schematically illustrated in Figure 6 where the SDS micellarization of individual and small SWNT bundles leads to controlled dissolution of the SWNT nanofibrils. Since these SWNT nanofibrils have not been annealed at this stage, the packing density of these nanotubes is expected to be reduced as a result of a number of terminal and sidewall defects as well as adsorbed species. These defects provide the necessary space for the SDS surfactants to adsorb on individual and small SWNT bundles to impart nanofibril dissolution. Figure 7a-d illustrate FESEM images of two SWNT nanofibrils before (a and b) and after (c and d, respectively) shortening with SDS solution. The ragged morphology at the tip end of these shortened nanofibrils (e and f, respectively) is attributed to their mosaic structures composed of multiple SWNTs of various lengths, as schematically shown in Figure 6.

Atomic force microscopy results indicate that these shortened nanofibrils are quite stiff after annealing at $10^{-6}$ torr vacuum and temperatures up to 450°C for 1hr (Nanoprobe B in Figure 7). Such vacuum annealing assists in densification of these nanofibrils by removing residual defects and adsorbates, leading to a more intimate rope-lattice packing. This also results in an enhanced resistance to SDS and other surfactant-assisted dissolution which can be beneficial to a number of electrochemical applications involving biological aqueous cultures. Figure 8 depicts the effective spring constant ($k_{eff}$) of these SWNT-nanofibril equipped AFM probes as a function of vacuum annealing temperature. For this, the SWNT nanofibrils were assembled onto soft levers through dielectrophoresis. The annealing induced a three fold improvement of stiffness. This is typically obtained of *ca.* 300°C and subsequent annealing at higher temperatures results minimum or no improvement in stiffness. At temperature of 450°C, the SWNT nanofibrils become brittle and tend to break upon repetitive indentation on a Si substrate. An FESEM image of a 450°C annealed fractured nanofibril is shown in Figure 8 (inset at lower right), containing a small stump at the end of the AFM probe.

**3.3.2 Spatial Functionalization of Nanoprobes**

As described in the introduction, AFM can be and has been used to investigate spatial electrochemistry and surface functionality of various samples. One of the applications of interest is using these shortened SWNT-nanofibril probes as nanoneedles to pierce a cell and establish a minimally invasive electrical communication with the inner biochemistry. Recently, Obataya *et al.*[12] have demonstrated the penetration of a living cell using a FIB-machined Si tip. With this application in mind, such electrochemical nanoneedle[10] should be surface functionalized with moieties exhibiting enhanced interaction with a variety of biological moieties. Figure S6 of Supporting Information illustrates two nanofibrils obtained by dielectrophoretic assembly from aqueous dispersion of covalently and non-covalently functionalized SWNTs. A DNA oligomer $(GT)_{20}$ (Figure S6a) was utilized for non-covalent solubilization of SWNTs.[48] Aqueous dispersion of SWNTs was also obtained by covalent functionalization onto carboxyl functionalities of oxidized SWNTs with flavin mononucleotide (FMN) moieties (Figure S6b).[49] FMN is a very important redox molecule in the electron transfer cycle of bacteria and cells thus it has been used extensively for biosensor applications.[50, 51] As shown before, functionalization of SWNTs with covalent chemistry typically increase their resistivity and render them unsuitable for long range conduction of electrons through these nanofibrils.[52] An ideal nanofibril configuration that overcomes this problem could involve hybrid structures of either linear or radial stacking of high and low conductivity nanotubes, as shown in Figure 9. For this, high intrinsic conductivity nanotubes would be attached onto the metal-coated AFM probe and preferably annealed to increase stiffness and solubility resistance. Subsequently a variety of functionalized SWNTs would be grown on either end or top of these nanofibrils imparting the desired functionality. This can be achieved by a second dielectrophoretic growth of functionalized SWNTs on top of previously grown nanofibrils.

Figure 10 illustrates the FESEM images of Type A heterostructure (longitudinal stacking) composed of a pristine and annealed SWNT-nanofibril (with diameter of *ca.* 100 nm) that has been extended with a much thinner (*ca.* 25 nm in diameter). As previously described, the pristine SWNT nanofibril was grown dielectrophoretically onto the AFM tip using original SWNT suspension, shortened with SDS solutions to the desired length (Figure 10a, b and c) and annealed at 300°C to impart greater stability against dissolution. Then the solution was replaced with FMN-SWNTs suspension and the tip of the previously grown nanofibril was immersed into the new solution at the desired depth to repeat the dielectrophoretic assembly. Figure 10d depicts that the 2$^{nd}$ dielectrophoretic growth resulted in the growth of a new nanofibril at the end of the previously grown one. As expected, the interface between the two nanofibrils shows a slight bump (Figure 10e) as a result of the partial overgrowth of FMN-functionalized SWNTs onto the initial nanofibril. The misorientation of the 2$^{nd}$ nanofibril was due to its thin nature that caused it to bend under intense heat from the electron-beam irradiation. This caused its free end to sway rapidly (see Figure S7 of the Supporting Information) under FESEM observation. The length of the 2$^{nd}$ grown nanofibril can also be trimmed to the desired length using the previously established SDS shortening procedure.

Figure 11 illustrates FESEM images of Type B (a–c) and Type C (d) nanoprobes, which are heterostructures in radial stacking and both radial and longitudinal stacking respectively. Figure 11a and its inset showed a shortened SWNT nanofibril used to overgrow the 2$^{nd}$ radial component (i.e. FMN functionalized SWNTs). Subsequent re-growth (first, second and third) takes place at the radial direction (Figure b–d, respectively) upon sequentially re-immersing the same nanofibril into FMN-functionalized SWNT suspension and repeating the dielectrophoretic assembly. The radial growth thickness can be controlled by varying immersion time and SWNT dispersion concentration. The pulling speed plays a lesser role due to the shorter length of the starting (inner) nanofibril. Figure 11b-d demonstrate close-ups of the thickened nanofibrils that increase by about 10 nm at each immersion/dielectrophoretic growth. In Figure 11d, besides radial thickening, longitudinal growth was also achieved by pulling past the end of the inner nanofibril. Such composite nanoprobes could provide several advantages: (1) The attachment of functional nanotubes in both radial and longitudinal direction maximizes the functionalization area with minimal decrease in conductivity; (2) The thickening in radial direction enhances nanoprobes' stiffness preventing them from potential bending or buckling; and (3) the growth of a relatively thinner terminal component enables high lateral resolution and could facilitate easier penetration into biological entities.

Polarization Raman measurements were employed to investigate the orientation of the re-grown SWNTs in the nanofibril in Figure 11c. The incident laser beam power was reduced to 1% (0.035 mW) to ensure the Raman signal originates mostly from the re-growth skin layer. As indicated in Figure S8, the obtained peak height ratio $I_0/I_{90}$ at the G band (1592 cm$^{-1}$) was *ca.* 4.6, comparable to that of the first grown nanofibril in Figure 11a. This indicates that re-growth maintains the high degree of nanotube alignment.

## 4. Conclusions

The positive dielectrophoretic organization of SWNT nanofibrils onto conductive AFM probes was investigated in order to impart better control over nanofibril dimensions, orientation and morphology. For this, several pertinent factors were investigated including nanotube dispersion concentration, immersion time, pulling rate, humidity and AFM probe alignment. After assembly, the length of the nanofibril can be further trimmed using SDS-assisted solubilization. Vacuum annealing these nanofibrils of 300°C increased their stiffness and reduced solubilization in SDS solution. Moreover, longitudinal and radial re-growth was achieved by re-immersion of shortened nanofibrils at the desired depth in a new SWNT suspension and repeating the dielectrophoretic process. These SWNT nanofibril AFM probes are mechanically robust and electrically conductive to act as structural nanoneedles and electrochemical nanoprobes. Such nanofibril-decorated scanning probes could find a number of applications in biomechanics and electrochemical analysis of biological entities as well as deep trench structures.

**Acknowledgement.** The authors would like to thank S. Kim for initial assistance with the experimental set-up for dielectrophoresis and Z. Luo for supplying DNA-wrapped SWNTs. Financial support from U.S. Army Research Office (grant # ARO-DAAD-19-02-1-0381), AFOSR (grant # FAJ550-06-1-0030) and NSF (grant # DMI 030395) are greatly appreciated.

**Supporting Information Available:** FESEM images of Olympus AC240 AFM probes used in the experiments (Figure S1); Photographs of the apparatus used for dielectrophoretic assembly of SWNT nanofibril-equipped AFM probes (Figure S2); AFM 3D images of a Si grid obtained with normal AC240 AFM probes and a SWNT nanofibril-equipped AFM probes (Figure S3); *I-V* curve of a SWNT nanofibril-equipped AFM probe (Figure S4); Comparison of the water evaporation rate in the open space and an environment-controlled cell (Figure S5); SWNT nanofibrils assembled from DNA-wrapped SWNTs and FMN-functionalized SWNTs (Figure S6); Rapid swaying of a thin FMN-functionalized SWNT nanofibril grown at the end of a much thicker pristine SWNT nanofibril under electron-beam exposure (Figure S7); Polarization dependent Raman spectra of the re-grown SWNT skin layer on nanotube nanofibrils (Figure S8) (PDF). This material is available free of charge via the Internet at http://pubs.acs.org.

REFERENCES

(1) Bonnell, D. A. *Scanning probe microscopy and spectroscopy: theory, techniques, and applications*; Wiley-VCH: New York, 2001.

(2) Frisbie, C. D.; Rozsnyai, L. F.; Noy, A.; Wrighton, M. S.; Lieber, C. M. *Science* **1994,** *265*, 2071.

(3) Noy, A.; Frisbie, C. D.; Rozsnyai, L. F.; Wrighton, M. S.; Lieber, C. M. *J. Am. Chem. Soc.* **1995,** *117*, 7943.

(4) Nguyen, C. V.; Stevens, R. M. D.; Barber, J.; Han, J.; Meyyappan, M.; Sanchez, M. I.; Larson, C.; Hinsberg, W. D. *Appl. Phys. Lett.* **2002,** *81*, 901.

(5) Cheung, C. L.; Hafner, J. H.; Lieber, C. M. *Proc. Natl. Acad. Sci. U. S. A.* **2000,** *97*, 3809.

(6) Hafner, J. H.; Cheung, C. L.; Woolley, A. T.; Lieber, C. M. *Prog. Biophys. Mol. Biol.* **2001,** *77*, 73.

(7) Nguyen, C. V.; Chao, K. J.; Stevens, R. M. D.; Delzeit, L.; Cassell, A.; Han, J.; Meyyappan, M. *Nanotechnol.* **2001,** *12*, 363.

(8) Yu, M.-F.; Files, B. S.; Arepalli, S.; Ruoff, R. S. *Phys. Rev. Lett.* **2000,** *84*, 5552.

(9) Wong, S. S.; Harper, J. D.; Lansbury, P. T.; Lieber, C. M. *J. Am. Chem. Soc.* **1998,** *120*, 603.

(10) Boo, H.; Jeong, R. A.; Park, S.; Kim, K. S.; An, K. H.; Lee, Y. H.; Han, J. H.; Kim, H. C.; Chung, T. D. *Anal. Chem.* **2006,** *78*, 617.

(11) Campbell, J. K.; Sun, L.; Crooks, R. M. *J. Am. Chem. Soc.* **1999,** *121*, 3779.

(12) Obataya, I.; Nakamura, C.; Han, S.; Nakamura, N.; Miyake, J. *Nano Lett.* **2005,** *5*, 27.

(13) Dai, H.; Hafner, J. H.; Rinzler, A. G.; Colbert, D. T.; Smalley, R. E. *Nature* **1996,** *384*, 147.

(14) Nishijima, H.; Kamo, S.; Akita, S.; Nakayama, Y.; Hohmura, K. I.; Yoshimura, S. H.; Takeyasu, K. *Appl. Phys. Lett.* **1999,** *74*, 4061.

(15) Stevens, R.; Nguyen, C.; Cassell, A.; Delzeit, L.; Meyyappan, M.; Han, J. *Appl. Phys. Lett.* **2000,** *77*, 3453.

(16) Akita, S.; Nishijima, H.; Nakayama, Y.; Tokumasu, F.; Takeyasu, K. *J. Phys. D: Appl. Phys.* **1999,** *32*, 1044.

(17) Hafner, J. H.; Cheung, C. L.; Oosterkamp, T. H.; Lieber, C. M. *J. Phys. Chem. B* **2001,** *105*, 743.

(18) Hafner, J. H.; Cheung, C. L.; Lieber, C. M. *Nature* **1999,** *398*, 761.

(19) Wade, L. A.; Shapiro, I. R.; Ma, Z. Y.; Quake, S. R.; Collier, C. P. *Nano Lett.* **2004,** *4*, 725.

(20) Zhang, J.; Tang, J.; Yang, G.; Qiu, Q.; Qin, L.-C.; Zhou, O. *Adv. Mater.* **2004,** *16*, 1219.

(21) Tang, J.; Yang, G.; Zhang, Q.; Parhat, A.; Maynor, B.; Liu, J.; Qin, L.-C.; Zhou, O. *Nano Lett.* **2005,** *5*, 11.

(22) Lee, H. W.; Kim, S. H.; Kwak, Y. K.; Han, C. S. *Rev. Sci. Instrum.* **2005,** *76*, 046108_1.

(23) Tang, J.; Gao, B.; Geng, H.; Velev, O. D.; Qin, L.-C.; Zhou, O. *Adv. Mater.* **2003,** *15*, 1352.

(24) Liu, J.; Rinzler, A. G.; Dai, H.; Hafner, J. H.; Bradley, R. K.; Boul, P. J.; Lu, A.; Iverson, T.; Shelimov, K.; Huffman, C. B.; Rodriguez-Macias, F.; Shon, Y.-S.; Lee, T. R.; Colbert, D. T.; Smalley, R. E. *Science* **1998,** *280*, 1253.

(25) Liu, J.; Casavant, M. J.; Cox, M.; Walters, D. A.; Boul, P.; Lu, W.; Rimberg, A. J.; Smith, K. A.; Colbert, D. T.; Smalley, R. E. *Chem. Phys. Lett.* **1999,** *303*, 125.

(26) Chattopadhyay, D.; Galeska, I.; Papadimitrakopoulos, F. *Carbon* **2002,** *40*, 985.

(27) Wei, H.; Kim, S. N.; Marcus, H. L.; Papadimitrakopoulos, F. *Chem. Mater.* **2006,** *18*, 1100.

(28) Hermanson, K. D.; Lumsdon, S. O.; Williams, J. P.; Kaler, E. W.; Velev, O. D. *Science* **2001,** *294*, 1082.

(29) Lumsdon, S. O.; Scott, D. M. *Langmuir* **2005,** *21*, 4874.

(30) Lumsdon, S. O.; Kaler, E. W.; Williams, J. P.; Velev, O. D. *Appl. Phys. Lett.* **2003,** *82*, 949.

(31) Zheng, L. F.; Brody, J. P.; Burke, P. J. *Biosens. Bioelectron.* **2004,** *20*, 606.

(32) Gray, D. S.; Tan, J. L.; Voldman, J.; Chen, C. S. *Biosens. Bioelectron.* **2004,** *19*, 1765.

(33) Lapizco-Encinas, B. H.; Simmons, B. A.; Cummings, E. B.; Fintschenko, Y. *Anal. Chem.* **2004,** *76*, 1571.

(34) Kim, J. E.; Han, C. S. *Nanotechnol.* **2005,** *16*, 2245.

(35) Seo, H. W.; Han, C. S.; Choi, D. G.; Kim, K. S.; Lee, Y. H. *Microelectron. Eng.* **2005,** *81*, 83.

(36) Krupke, R.; Hennrich, F.; von Lohneysen, H.; Kappes, M. M. *Science* **2003,** *301*, 344.

(37) Krupke, R.; Hennrich, F.; Weber, H. B.; Kappes, M. M.; von Lohneysen, H. *Nano Lett.* **2003,** *3*, 1019.

(38) Duesberg, G. S.; Loa, I.; Burghard, M.; Syassen, K.; Roth, S. *Phys. Rev. Lett.* **2000,** *85*, 5436.

(39) Gommans, H. H.; Alldredge, J. W.; Tashiro, H.; Park, J.; Magnuson, J.; Rinzler, A. G. *J. Appl. Phys.* **2000,** *88*, 2509.

(40) Chattopadhyay, D.; Lastella, S.; Kim, S.; Papadimitrakopoulos, F. *J. Am. Chem. Soc.* **2002,** *124*, 728.

(41) Esplandiu, M. J.; Bittner, V. G.; Giapis, K. P.; Collier, C. P. *Nano Lett.* **2004,** *4*, 1717.

(42) Burt, D. P.; Wilson, N. R.; Weaver, J. M. R.; Dobson, P. S.; Macpherson, J. V. *Nano Lett.* **2005,** *5*, 639.

(43) Martinez, J.; Yuzvinsky, T. D.; Fennimore, A. M.; Zettl, A.; Garcia, R.; Bustamante, C. *Nanotechnol.* **2005,** *16*, 2493.

(44) Li, L. J.; Nicolas, R. J.; Chen, C. Y.; Darton, R. C.; Baker, S. C. *Nanotechnol.* **2005,** *16*, S202.

(45) Richard, C.; Balavoine, F.; Schultz, P.; Ebbesen, T. W.; Mioskowski, C. *Science* **2003,** *300*, 775.

(46) O'Connell, M. J.; Bachilo, S. M.; Huffman, C. B.; Moore, V. C.; Strano, M. S.; Haroz, E. H.; Rialon, K. L.; Boul, P. J.; Noon, W. H.; Kittrell, C.; Ma, J. P.; Hauge, R. H.; Weisman, R. B.; Smalley, R. E. *Science* **2002,** *297*, 593.

(47) Wanless, E. J.; Ducker, W. A. *J. Phys. Chem.* **1996,** *100*, 3207.

(48) Zheng, M.; Jagota, A.; Semke, E. D.; Diner, B. A.; Mclean, R. S.; Lustig, S. R.; Richardson, R. E.; Tassi, N. G. *Nat. Mater.* **2003,** *2*, 338.

(49) Ju, S.-Y.; Papadimitrakopoulos, F. Manuscript in preparation.

(50) Katz, E.; Willner, I. *ChemPhysChem* **2004,** *5*, 1084.

(51) Wang, J. *Electroanalysis* **2005,** *17*, 7.

(52) Bahr, J. L.; Tour, J. M. *J. Mater. Chem.* **2002,** *12*, 1952.

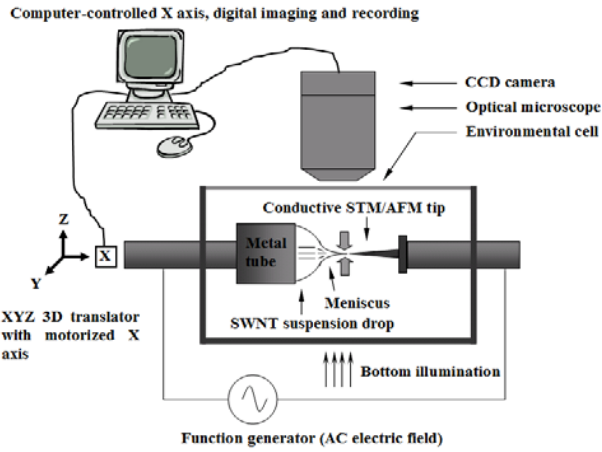

**Scheme 1.** Schematic representation of the experimental set-up for fabricating SWNT AFM probes via dielectrophoretic assembly.

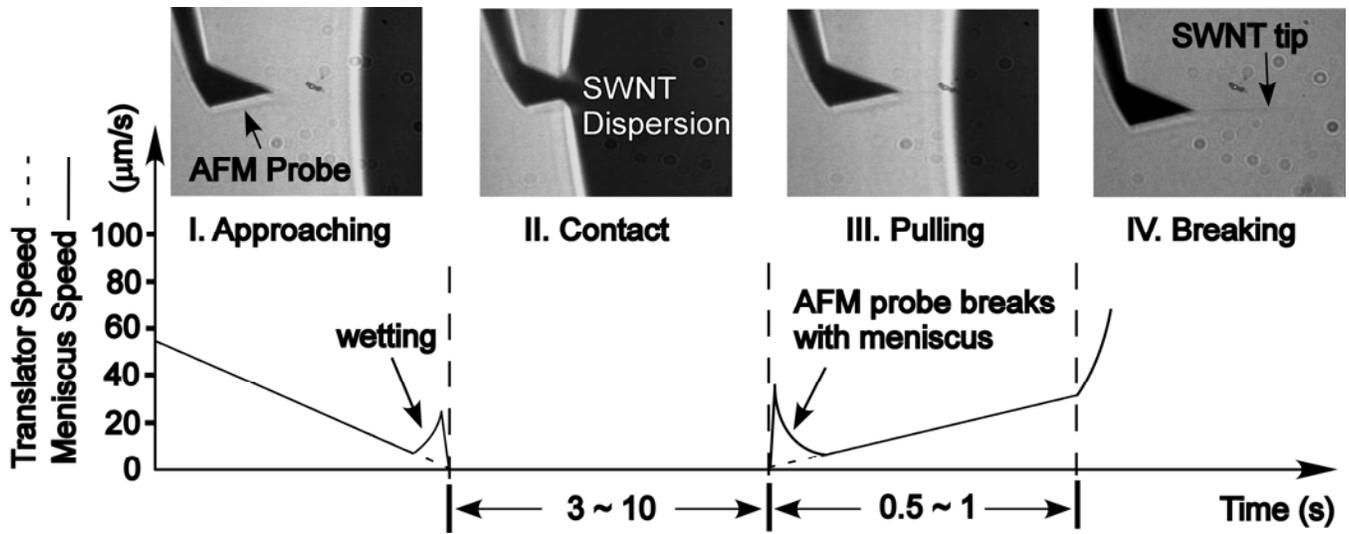

**Figure 1.** Schematic illustration of assembling SWNT nanofibrils onto AFM probes via dielectrophoresis.

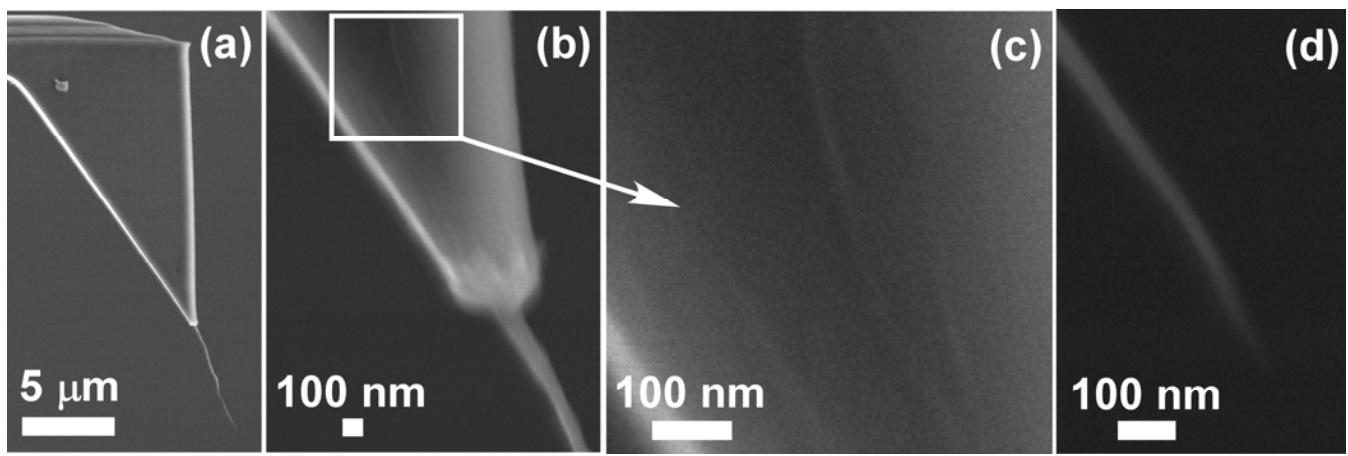

**Figure 2.** Representative FESEM images of a SWNT nanofibril grown dielectrophoretically onto an AFM probe. Overall view (a), interface region between the AFM probe and the SWNT nanofibril (b), representative SWNT bundles adhered onto the sidewall of the AFM probe surface with direction parallel to the pulling axis (c) and nanofibril end consisting of coalesced SWNTs that result in a sharp end (d).

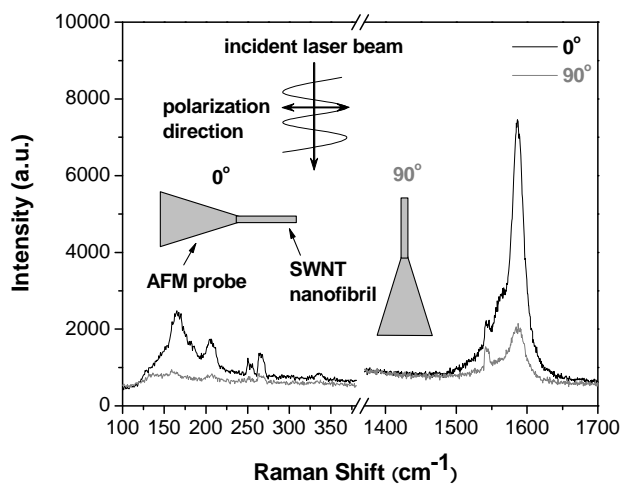
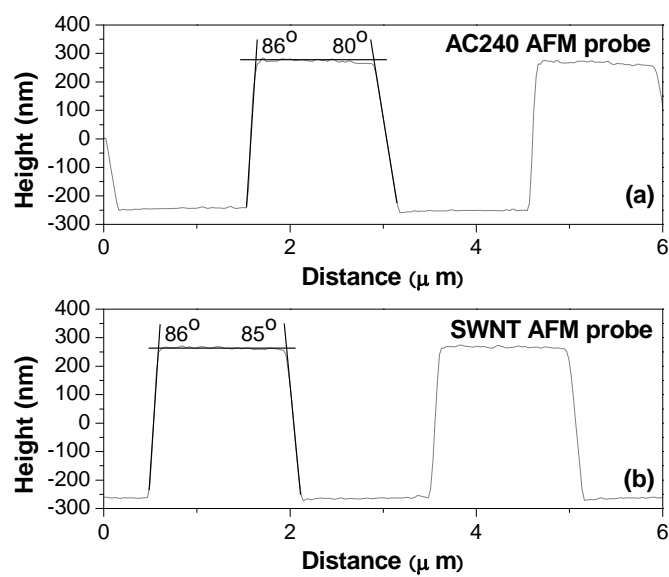

**Figure 3.** Polarized Raman spectra of SWNT nanofibrils dielectrophoretically grown onto AFM probes. Inset illustrates the experimental geometry with respect to the incident 785 nm polarized laser beam.

**Figure 4.** Scanning probe microscopy height profile of a Si grid measured with a typical AFM probe (AC240) (a), and a SWNT-nanofibril equipped high-aspect-ratio AFM probe (b) (see text for details).

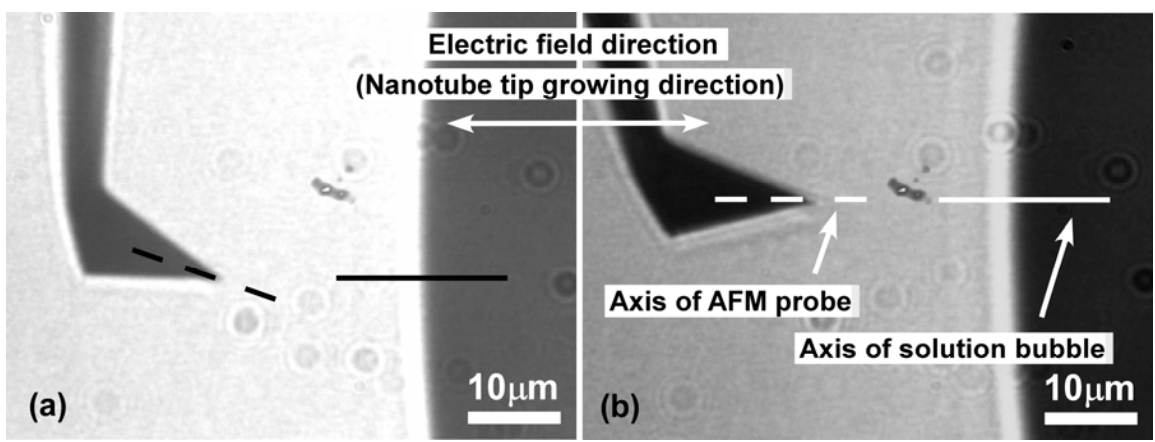

**Figure 5.** Optical images of the placement of an AC240 AFM probe on a platform with a tilting angle of 0° (a) and 26° (b). This 26° tilting results in a SWNT nanofibril grown parallel to the central axis of the AFM cone.

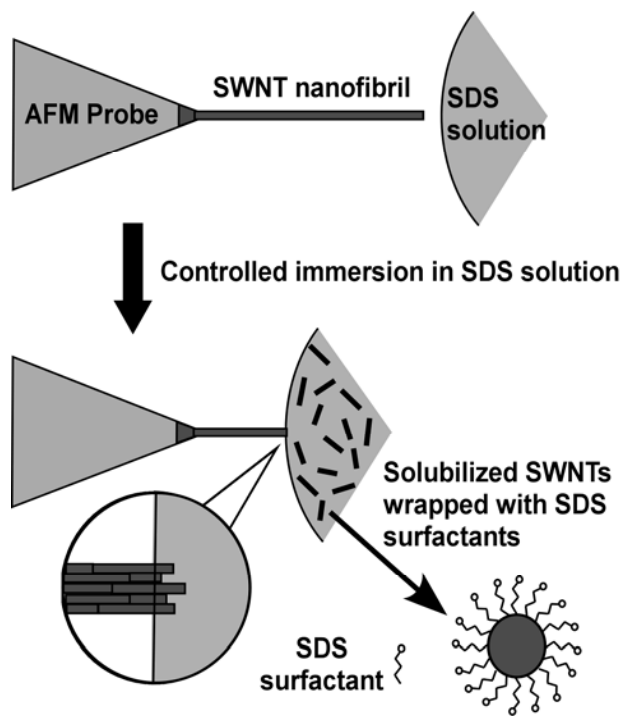
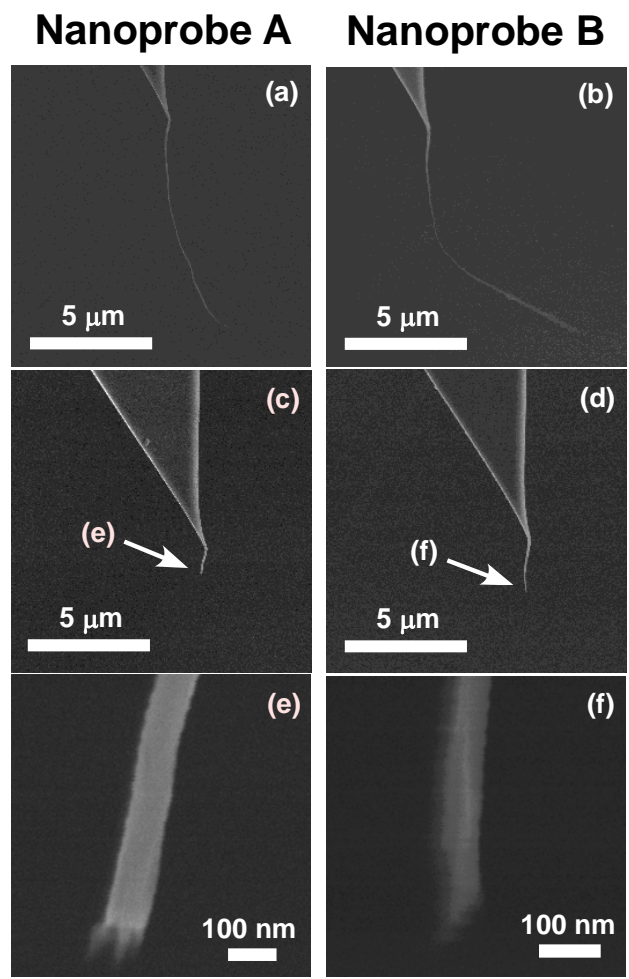

**Figure 6.** SDS-assisted shortening of SWNT nanofibrils through gradual solubilization upon immersion in an aqueous solution of sodium dodecyl sulfate (SDS).

**Figure 7.** FESEM images of two typical SWNT nanofibrils before (a and b) and after (c and d, respectively) upon immersion in an aqueous solution of SDS (1 wt%). The ragged ends of the shortened nanofibrils (e and f, respectively) are consistent with a mosaic nanotube structure of various lengths. After shortening, nanoprobe B (right column) was annealed at 450°C in vacuum ($10^{-6}$ torr) for 1hr.

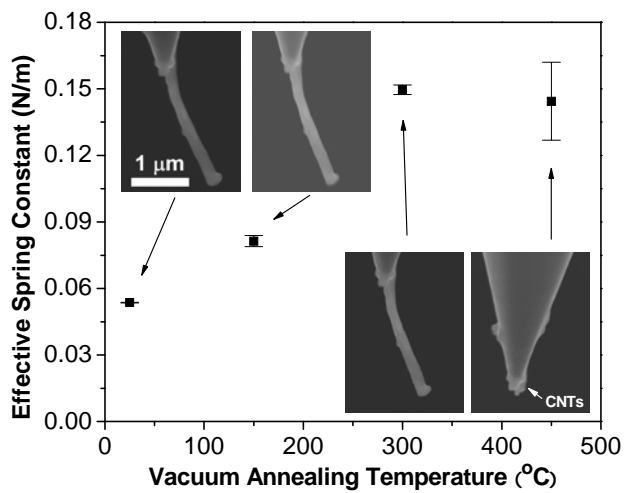

**Figure 8.** Effective spring constant of SWNT nanofibril equipped AFM probes as a function of annealing temperature at vacuum of $10^{-6}$ torr for 1 hr. Insets illustrate the FESEM images of these nanofibrils after mechanical indentation to determine their effective spring constant.

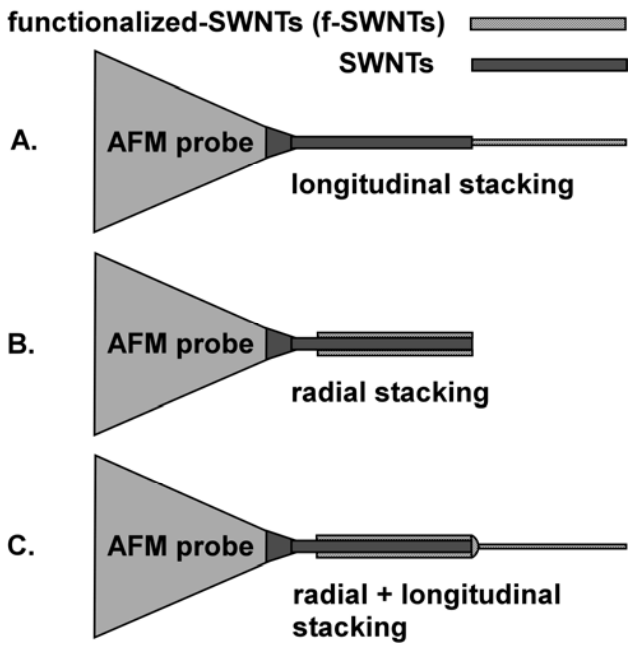
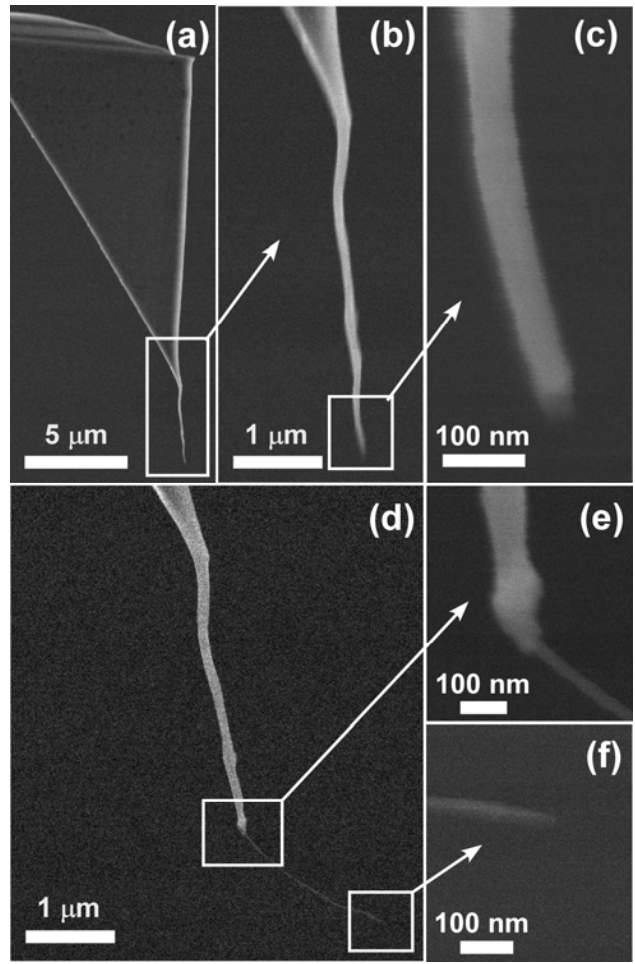

**Figure 9.** Schematic representation of heterostructure nanofibril configuration containing functionalized (f-) and pristine SWNTs.

**Figure 10.** FESEM images of a linear nanofibril heterostructure containing pristine and annealed (300°C) nanotubes (near the AFM probe) (a-c), which is subsequently extended with a FMN-functionalized SWNT nanofibril (d-f).

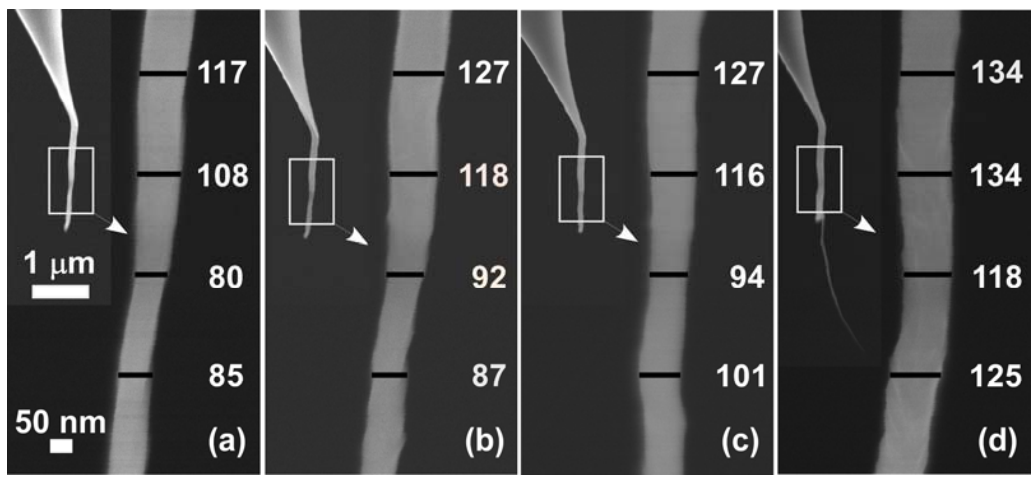

**Figure 11.** Type B (a-c) and Type C (d) nanoprobe heterostructures achieved by sequential dielectrophoretic growth of FMN-functionalized SWNTs on top of the previous nanofibril, starting with (a) and finishing with (d). The scale is 1 μm for the insets on top left and 50 nm for the close-ups. The nanofibril diameters (numbers in white) were measured at approximately the same sites after each growth. Longitudinal growth (d) was also achieved on (c) by pulling this nanofibril past the end of (c).

Table of Contents Synopsis

SWNT nanofibrils were dielectrophoretically assembled onto conductive AFM probes. The precise length trimming post-assembly was achieved through solubilization using SDS surfactants. Re-growth of a new functionalized SWNT nanofibril from the side or end of a previously grown nanofibril was demonstrated by sequential dielectrophoresis.

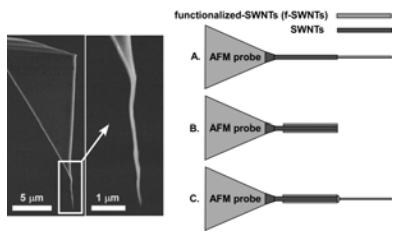

# Supporting information

# Control of Length and Spatial Functionality of Single-Wall Carbon Nanotube AFM Nanoprobes

*Haoyan Wei†, Sang Nyon Kim‡, Minhua Zhao†, Sang-Yong Ju‡, Bryan D. Huey†, Harris L. Marcus†\**

*and Fotios Papadimitrakopoulos‡\**

† Materials Science and Engineering Program, Department of Chemical, Materials and Biomolecular Engineering, Institute of Materials Science, University of Connecticut, Storrs, CT 06269, USA

‡ Nanomaterials Optoelectronics Laboratory, Polymer Program, Institute of Materials Science, Department of Chemistry, University of Connecticut, Storrs, CT 06269, USA

\* To whom correspondence should be addressed.

E-mail: hmarcus@mail.ims.uconn.edu, papadim@mail.ims.uconn.edu.

Fotios Papadimitrakopoulos: Phone: (860) 486-3447. Fax: (860) 486-4745.

Harris L. Marcus: Phone: (860) 486-4623. Fax: (860) 486-4745.

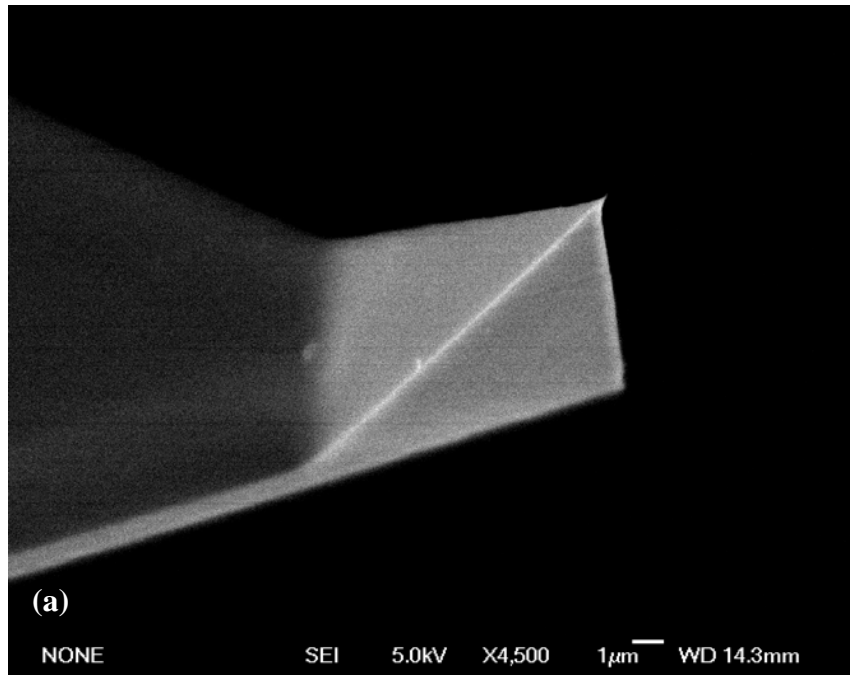
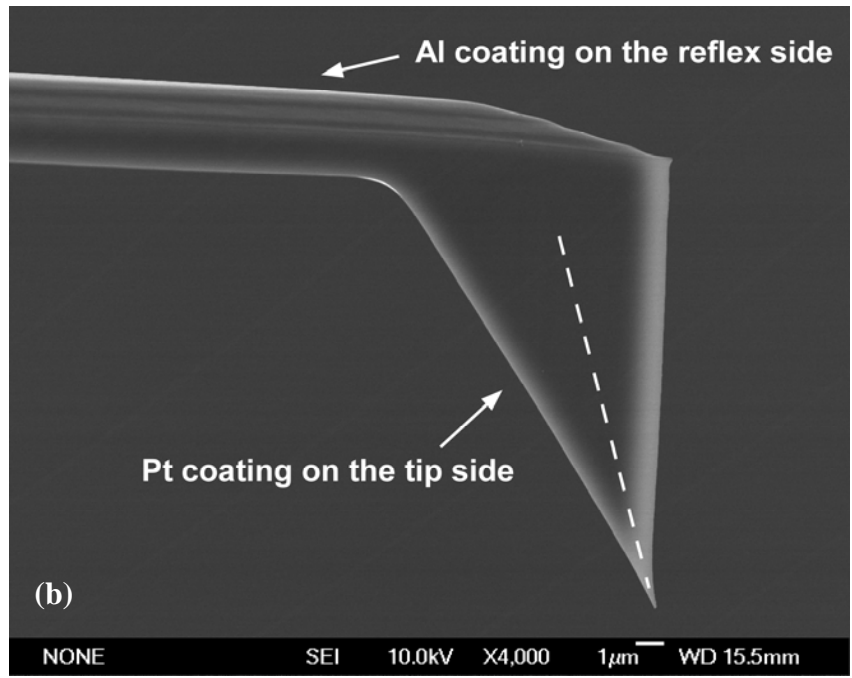

**Figure S1.** FESEM images of Olympus AC240 AFM probes. Oblique view from tip side (a) and horizontal side view (b). Central axis of the probe was indicated with a white dashed line in (b).

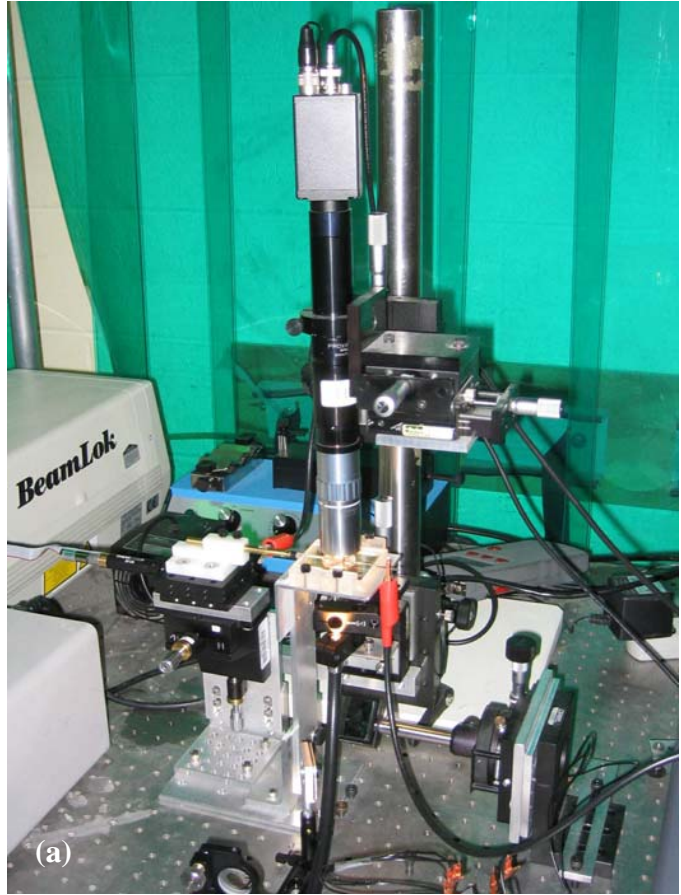

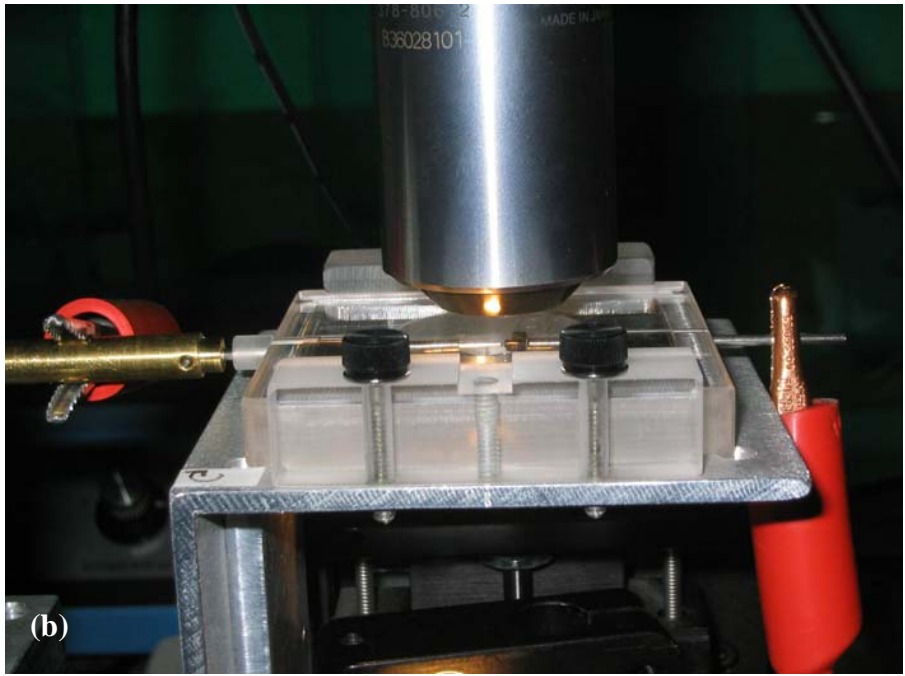

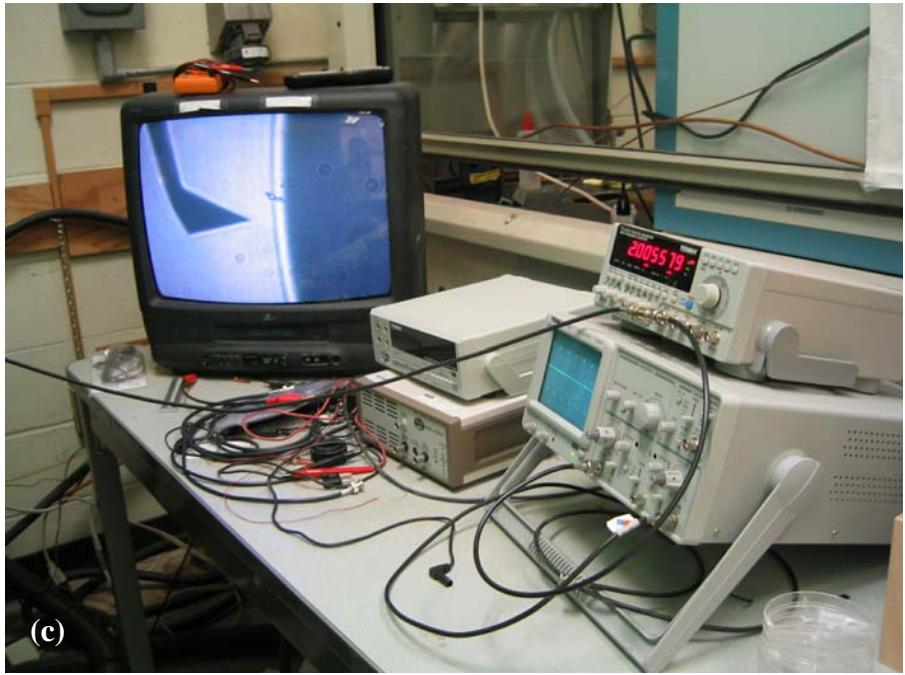

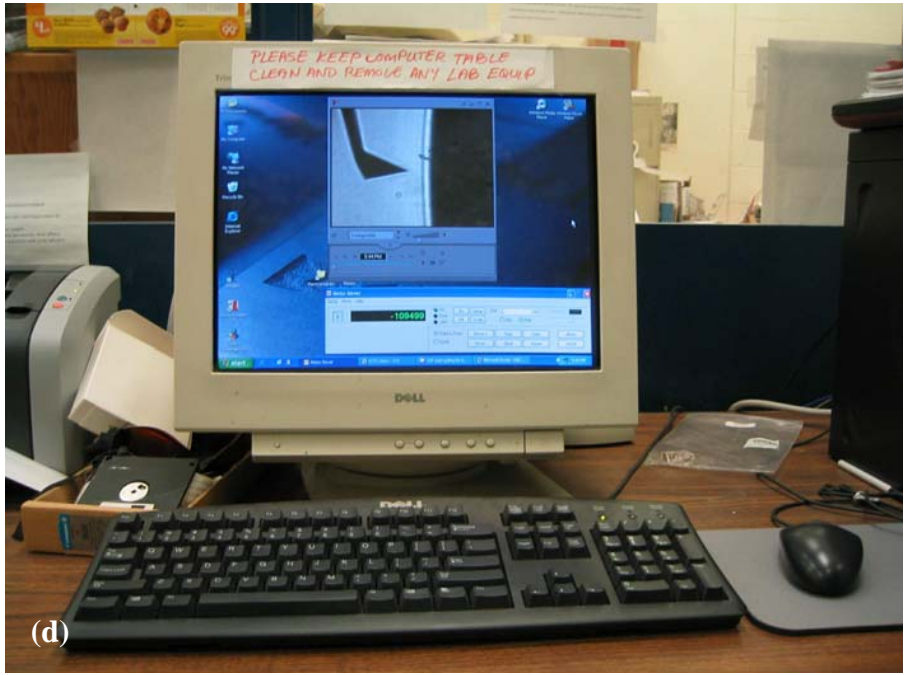

**Figure S2.** Photographs of the apparatus used for assembling SWNT nanofibrils onto AFM probes via dielectrophoretic process. Optical microscope, 3D translation stage and environmental cell (a), close-up view of environmental cell (b), TV monitor, function generator and oscilloscope (c), and computer station (d).

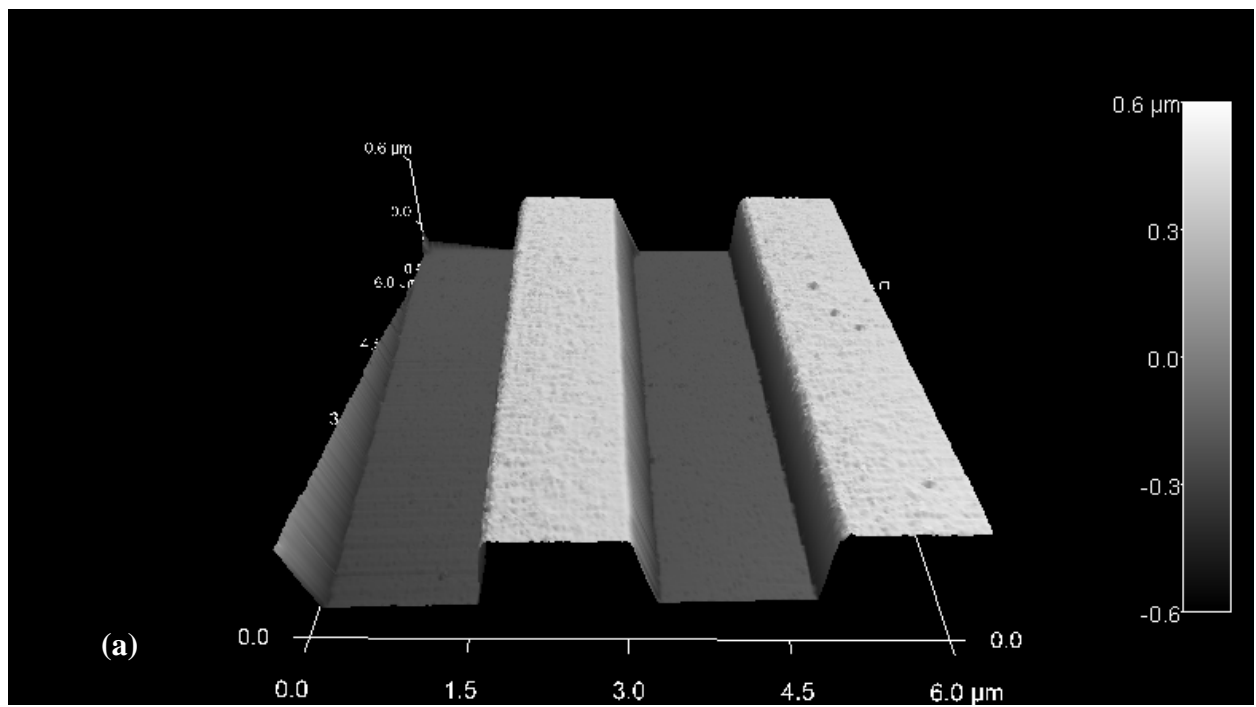

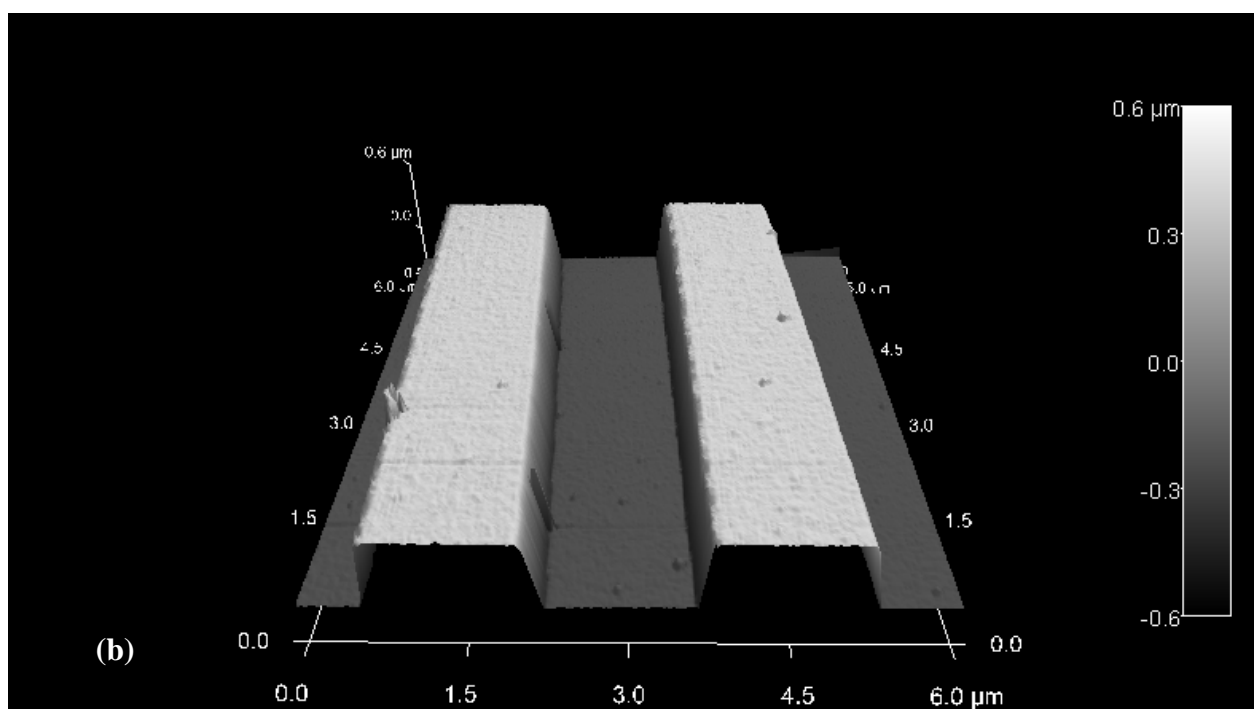

**Figure S3.** AFM 3D images of a Si grid obtained with a normal AC240 AFM probe (a) and a SWNT nanofibril-equipped AFM probe (b) respectively (see main text for details). The cross-sectional line profiles were shown in Figure 4 in the main text.

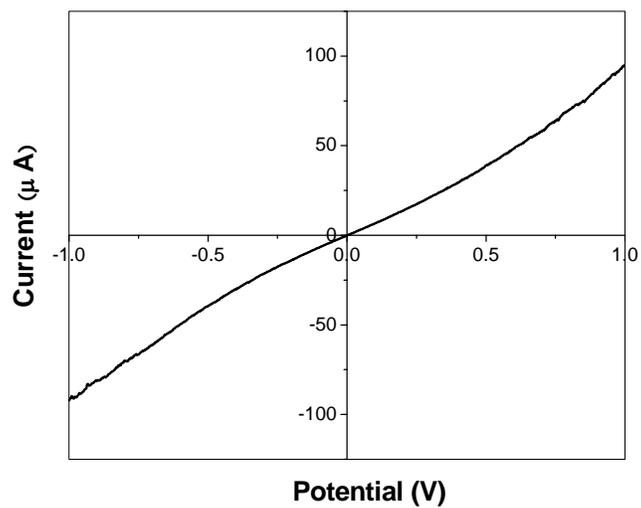

**Figure S4.** *I-V* curve of a SWNT nanofibril-equipped AFM probe in contact with a Pt substrate.

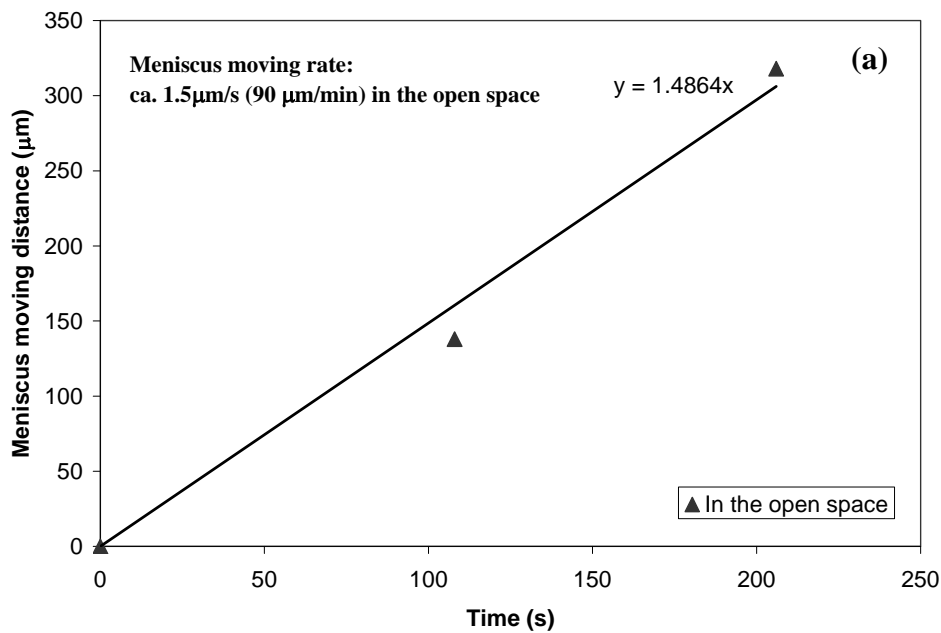

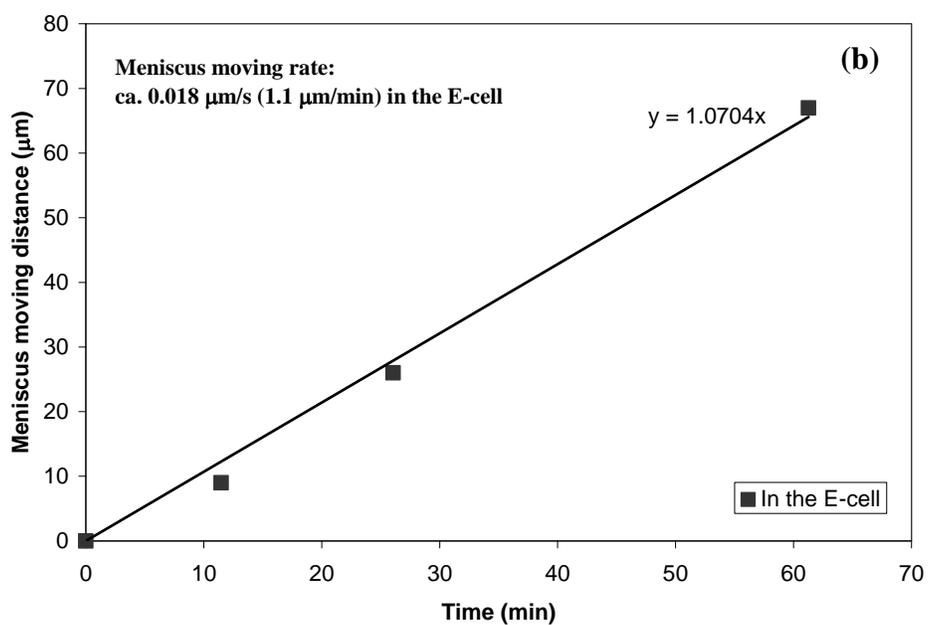

**Figure S5** Comparison of the water evaporation rate in the open space (a) and environment-controlled cell (b) using the meniscus moving rate as the indicator (Note: different time scales for X axis). The sealing slowed down the evaporation about 80 times.

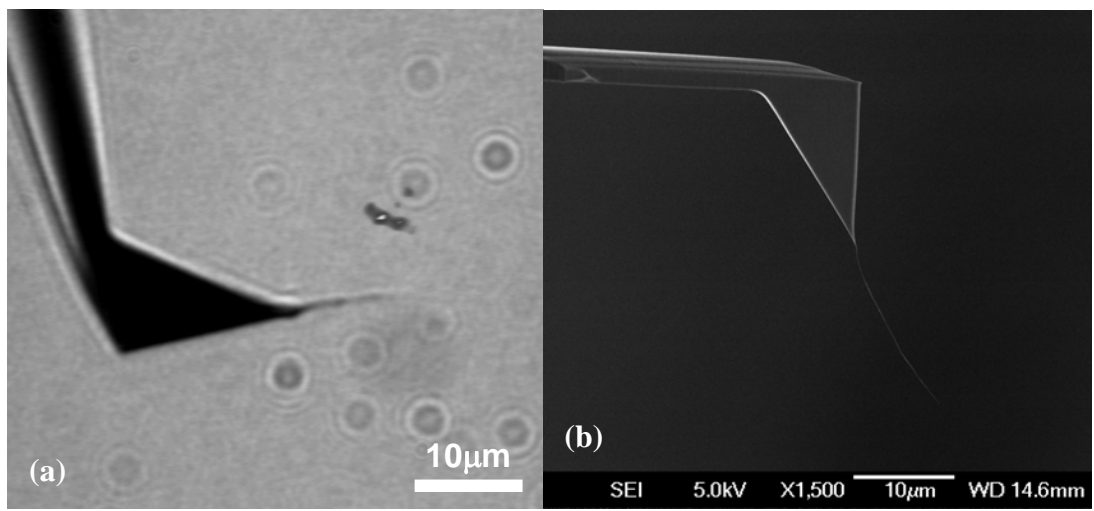

**Figure S6.** SWNT nanofibrils assembled via dielectrophoresis using DNA-wrapped (a) and FMN-functionalized (b) SWNTs dispersion.

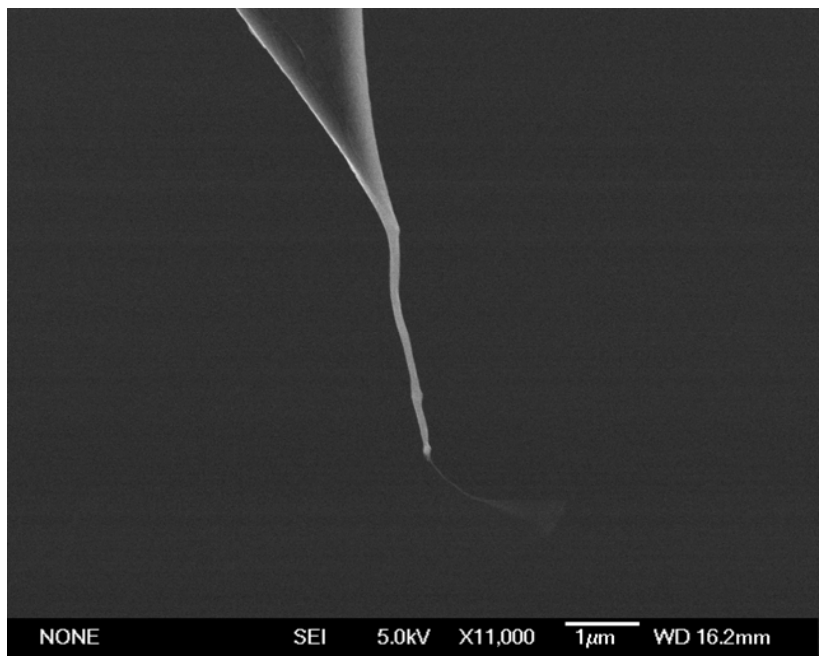

**Figure S7.** The thin FMN-functionalized SWNT nanofibril grown at the end of a much thicker pristine SWNT nanofibril displayed rapid swaying. The swaying amplitude increased with increasing electron beam exposure.

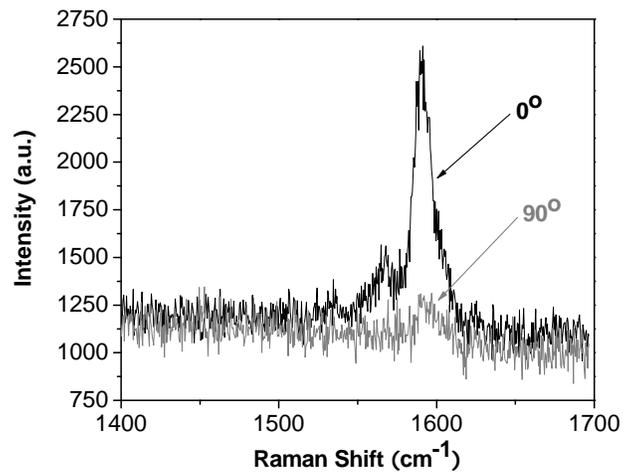

**Figure S8.** Polarization dependent Raman of Type B SWNT nanofibrils at reduced excitation power (1%) to ensure interrogation of the overcoating SWNT layer. The orientation function is comparable to that of Figure 3 in the main text.